\documentclass[journal]{IEEEtran}
\usepackage{amsmath,amsfonts}
\usepackage{algorithmic}
\usepackage{array}
\usepackage[caption=false,font=normalsize,labelfont=sf,textfont=sf]{subfig}
\usepackage{textcomp}
\usepackage{stfloats}
\usepackage{url}
\usepackage{verbatim}
\usepackage{graphicx}
\usepackage{cite}
\def\BibTeX{{\rm B\kern-.05em{\sc i\kern-.025em b}\kern-.08em
    T\kern-.1667em\lower.7ex\hbox{E}\kern-.125emX}}
\usepackage{balance}

\usepackage{booktabs}
\usepackage{threeparttable}
\usepackage{tabularx}
\usepackage{multirow}

\newcommand{\softmax}{so\!f\!tmax}
\newcommand{\TransEnc}{T\hspace{-0.075em}r\hspace{-0.05em}an\hspace{-0.05em}s\hspace{-0.075em}Enc}
\newcommand{\MHAtt}{M\!H\!Att}
\newcommand{\Head}{H\!ead}
\newcommand{\FFL}{F\!F\!L}

\newcommand{\BERT}{B\!E\!RT}

\newcommand{\cafe}{Caf\'{e}}
\newcommand{\Xone}{X\hspace{-0.05em}1\hspace{0.05em}}
\newcommand{\Xtwo}{X\hspace{-0.05em}2\hspace{0.05em}}
\newcommand{\ccol}[2]{ \multicolumn{#1}{c}{#2}}
\newcommand{\mydots}{\cdot\hspace{-0.2em}\cdot\hspace{-0.2em}\cdot}
\newcommand{\RQ}[2]{\vspace{0.5em}\noindent {\bf H#1} #2\vspace{0.5em}}

\begin{document}
\title{An Efficient Multimodal Learning Framework\\
to Comprehend Consumer Preferences \\
Using BERT and Cross-Attention}

\author{{
\Large Junichiro Niimi}$^{1, 2}$\\
(jniimi@meijo-u.ac.jp)\\
~\\
1. Faculty of Business Management, Meijo University\\
2. RIKEN Center for Advanced Intelligence Project (AIP)
}

\markboth{This manuscript has been submitted to the journal and is under peer review}%
{Niimi: An Efficient Multimodal Learning Framework to Comprehend Consumer Preferences Using BERT and Cross-Attention}

\maketitle

\begin{abstract}
Today, the acquisition of various behavioral log data has enabled deeper understanding of customer preferences and future behaviors in the marketing field. In particular, multimodal deep learning has achieved highly accurate predictions by combining multiple types of data. Many of these studies utilize with feature fusion to construct multimodal models, which combines extracted representations from each modality. However, since feature fusion treats information from each modality equally, it is difficult to perform flexible analysis such as the attention mechanism that has been used extensively in recent years. Therefore, this study proposes a context-aware multimodal deep learning model that combines Bidirectional Encoder Representations from Transformers (BERT) and cross-attention Transformer, which dynamically changes the attention of deep-contextualized word representations based on background information such as consumer demographic and lifestyle variables. We conduct a comprehensive analysis and demonstrate the effectiveness of our model by comparing it with six reference models in three categories using behavioral logs stored on an online platform. In addition, we present an efficient multimodal learning method by comparing the learning efficiency depending on the optimizers and the prediction accuracy depending on the number of tokens in the text data.
\end{abstract}


\section{Introduction}
Nowadays, social media and other online platforms play an important role in shaping consumer behaviors and aiding decision-making. However, amidst the burgeoning amount of online information, users often face difficulties in discovering preferred content and suitable services \cite{deeplearningRecommender}. To address this information overload, recommender systems have recently found application not only in social networking services (SNSs) and electronic commerce (EC) but also in wider domains such as tourism, healthcare, and education \cite{recommender_survey}. To optimize personalized content for each user, these systems must accurately discern the preferences of consumers with various sets of values to offer tailored recommendations.

With the evolution of machine learning techniques, contemporary models can handle a wide array of data, including text. Notably, Transformer \cite{transformer} has made substantial contributions to the field of natural language processing (NLP). Bidirectional Encoder Representations from Transformers (BERT) \cite{bert} is known as one of the significant models in this regard. Leveraging large language models (LLMs), BERT enables the prediction and classification of consumers based on the texts they contribute to the platform. Furthermore, many recommender systems leverage review texts posted on platforms, \cite{bert_hotel}, commonly referred to as electronic word-of-mouth (eWOM).

In addition, multimodal learning, which combines multiple types of data to derive joint representations for classification and regression tasks, has gained widespread adoption. In particular, simultaneous analysis of data such as text and images, previously difficult to analyze individually, is now being undertaken in conjunction with other modalities. Nonetheless, despite these advancements, multimodal learning in marketing studies remains relatively limited, primarily due to the specificity of the data and marketing-specific issues, notably consumer heterogeneity \cite{niimi_arxiv}.

Both the development of an optimal recommender system and the acquisition of review data are crucial on online platforms; however, the use of these data remains limited despite their potential value in understanding customer preferences. Therefore, in this study, we construct a novel multimodal deep learning model to assess user preferences on social platforms. The paper is structured as follows: First, we review prior studies relevant to our research. Next, we formulate hypotheses to address our research question. Subsequently, we outline the model architecture and provide an overview of the dataset. Then, we conduct several analyses to demonstrate the performance of the proposed model. Finally, we summarize the results and discuss the implications and challenges of our study.

\section{Related Study}
\subsection{Attention Mechanism}
\begin{figure}[t]
   \centering
\subfloat[][Attention mechanism]{\includegraphics[width=0.24\linewidth]{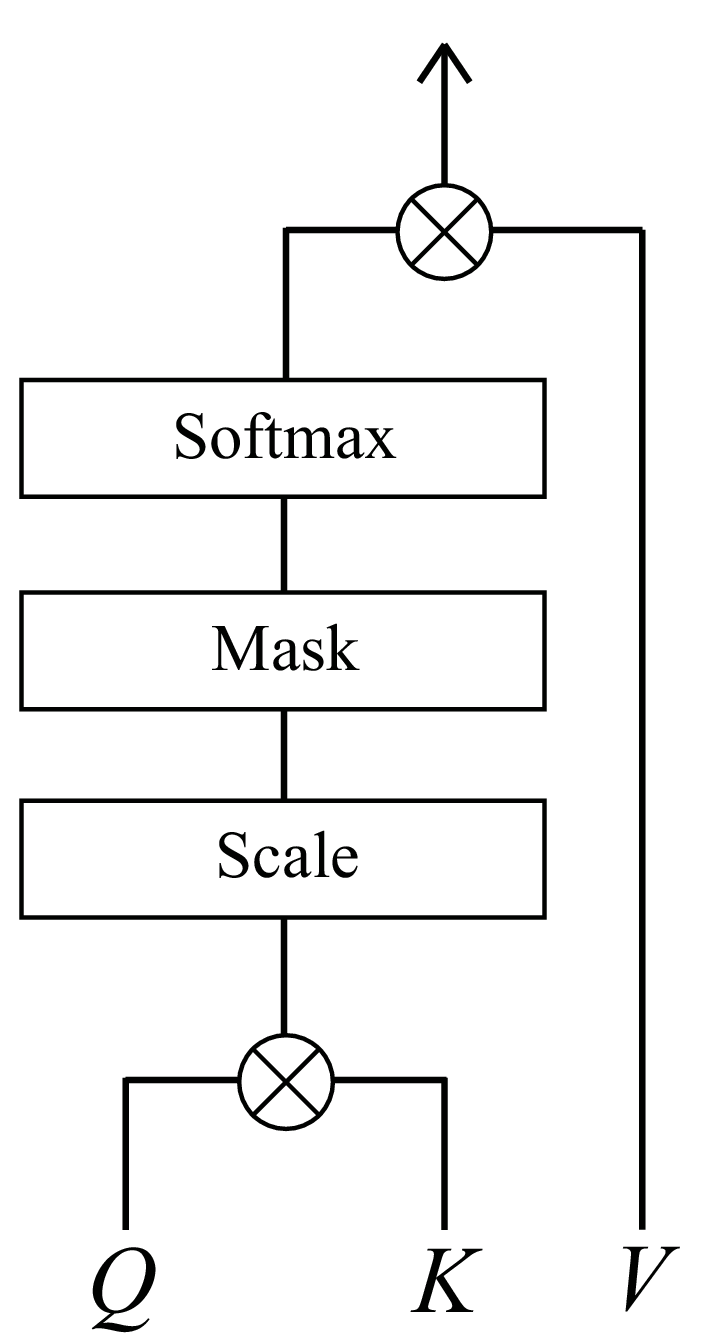}\label{fig:architecture_att}} 
\quad
\subfloat[][Multihead attention]{\includegraphics[width=0.29\linewidth]{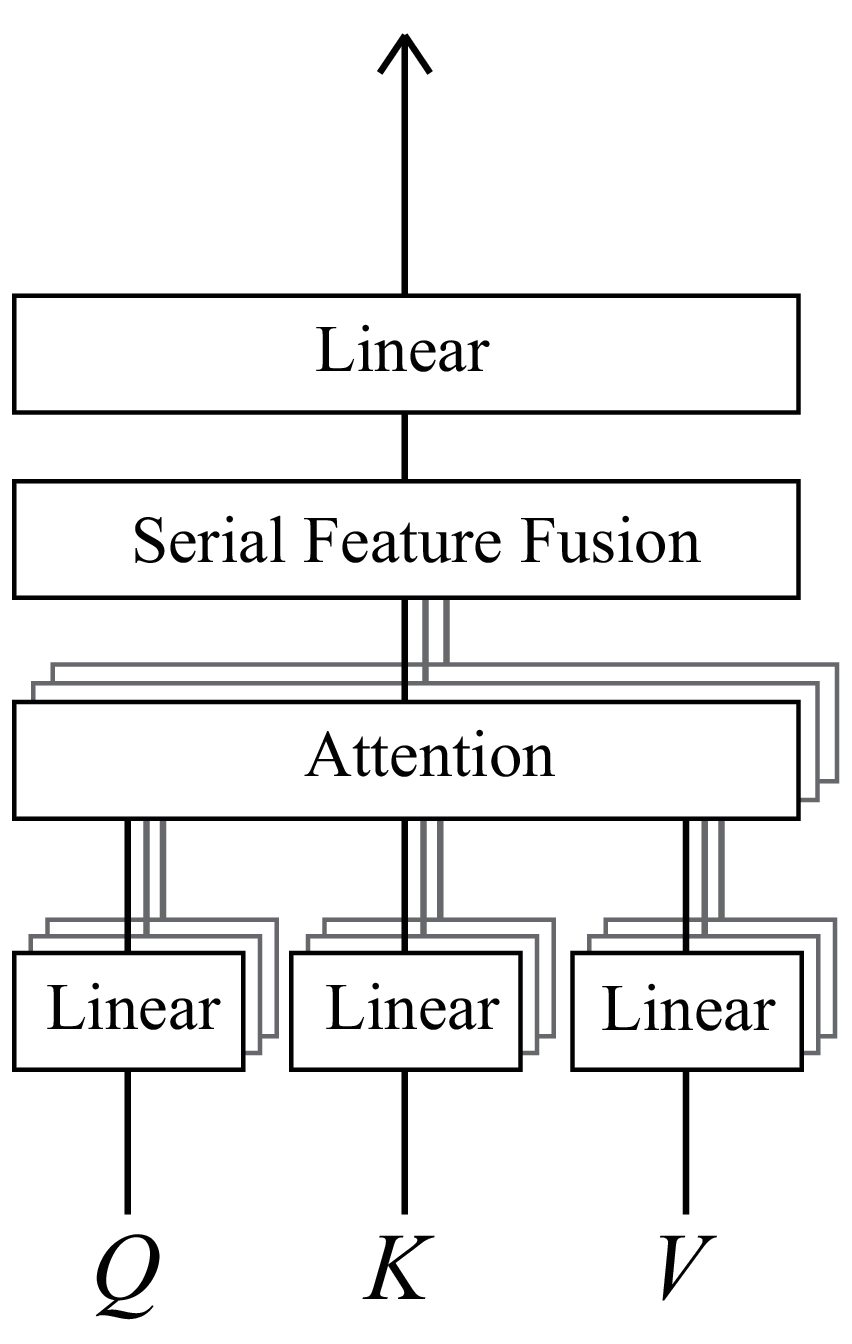}\label{fig:architecture_mhatt}} 
\quad
\subfloat[][Transformer encoder]{\includegraphics[width=0.28\linewidth]{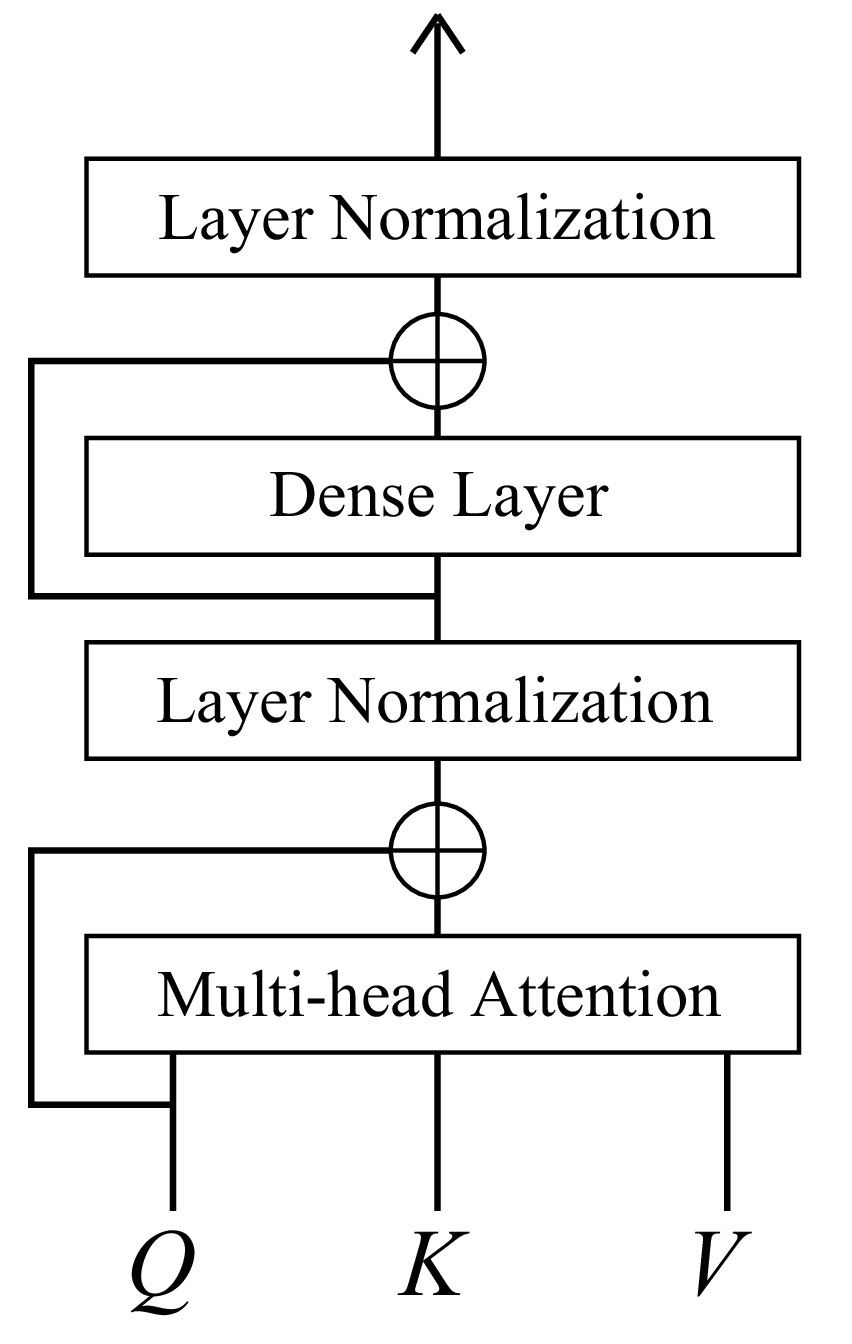}\label{fig:architecture_transenc}}
   \caption{From Attention to Transformer}
   \label{fig:architecture}
\end{figure}

First, it is essential to discuss the attention mechanism (Fig. \ref{fig:architecture_att}) \cite{attention}, which has had a significant impact on the field of machine learning. This mechanism operates by selectively focusing on relevant parts of the input sequence, thereby enabling models to prioritize and process these significant elements with greater emphasis. For example, a scaled-dot attention ($Att$) is computed using query ($Q$), key ($K$), value ($V$), and the softmax function ($\softmax$) as follows:
\begin{align}
   Att(Q, K, V) = \softmax \! \left(\frac{QK^T}{\sqrt{d_K}}\right)\!V
\end{align}

It adjusts the focus by computing attention weights within the softmax function, assigning relative importance to each element within the sequence. Particularly advantageous in handling large representations, attention can effectively train the model through layer-wise concatenation of multiple representations \cite{multimodal_att}. Two widely recognized variations of this mechanism are self-attention (SA) and source--target attention (STA), both obtained through the same calculation.
In terms of differences, SA involves query, key, and value for the source and computes relationships between elements within the source sequence. On the other hand, STA uses query and key for the source, and value for the target, computing relationships between the source and the target. For example, in the field of NLP, SA is utilized to identify word-to-word relationships within a sentence, thus providing contextual understanding.

\subsection{Transformer}
Based on the attention mechanism, Transformer distributes multiple attentions with weighted $Q$, $K$, and $V$ in parallel, a concept known as multihead attention ($\MHAtt$, see Fig. \ref{fig:architecture_mhatt}), which is expressed with $m$-th attention head ($\Head_m$, where $m = 1, 2, \ldots, M$) and $Att$ as \cite{transformer}:
\begin{align}
   \Head_m (Q,K,V) =& Att(QW_m^Q, ~KW_m^K, ~VW_m^V) \\
   \MHAtt(Q,K,V) =& concat(\Head_1, ~\Head_2, \mydots, \\
   &\Head_M) W^O
\end{align}
Transformer consists of an encoder and a decoder. Encoder's output ($\TransEnc$, see Fig. \ref{fig:architecture_transenc}) is obtained using layer-normalization ($LN$) \cite{layernorm}, feed-forward layer ($\FFL$), residual network \cite{resnet}, and $\MHAtt$ as:
\begin{align}
   \TransEnc &(Q,K,V) = LN(u+\FFL(u)) \\
   \text{where}~&u = LN(Q + \MHAtt (Q,K,V))
\end{align}
While prior research has proposed integrating the attention mechanism into recurrent models \cite{attention}, prior studies have shown that a single Transformer outperforms the combination of attention and recurrent structures \cite{attention_review}.

Whether discussing the attention mechanism or Transformer, some studies \cite{alaraj, niimi_jjas} have highlighted the utility of STA in capturing contextual information (i.e., the background) of the sequential data. Specifically, by setting tabular data (including demographic information) as the target, STA dynamically weighs the attention given to time-series data (including user behavior) as the source, based on demographic and other tabular variables.
In addition, other study \cite{crossAttentionDisaster} highlighted the utility of cross-attention of Transformer which integrates both visual and textual post about the same event on social media to evaluate whether the post is informative or not.

\subsection{BERT}
BERT stands out as one of the most significant pre-trained language models, which consists of Transformer encoder \cite{bert}. In the NLP field, the problem of ambiguity, where the meaning of a word changes depending on context, has long been recognized when handling textual data \cite{elmo}. Within the BERT architecture, SA plays an important role in obtaining distributed representations of textual data known as deep-contextualized word representations. This mechanism overcomes the ambiguity problem by adjusting embeddings based on context, i.e., the word's relationship to other words in the sentence, unlike traditional word-embedding methods such as word2vec \cite{word2vec}, which assign context-independent unique vectors \cite{bert, bert_googleplay}.

BERT utilizes fixed-length tokenization with padding and truncation. The full output shape of BERT is a 3-dimensional tensor with dimensions ($bs$, $len_{max}$, $param_{\BERT}$), where $len_{max}$ represents the aligned length of the tokenized sequence, and $param_{\BERT}$ depends on the scale of the BERT model (e.g., 768 for BERT-Base and 1024 for BERT-Large).
In addition, BERT has a pooler-output which is the 2-dimensional tensor with shape ($bs$, $param_{\BERT}$) obtained by applying a $tanh$ activation to a weighted sum of the [CLS] token. 
Pooler-output has been adopted in many downstream tasks due to its simplicity, effectively addressing ambiguity in natural language. For example, in marketing applications, a study \cite{bert_googleplay} utilized BERT to obtain deep-contextualized word representations from review text about mobile applications on online platforms, enabling the prediction of user loyalty.

In addition, various models based on BERT have been proposed such as a robustly optimized BERT pre-training approach (RoBERTa) \cite{roberta} and DistilBERT \cite{distilbert}. In particular, RoBERTa \cite{roberta} is the improved model of BERT, which performance through pre-training on a larger dataset and longer training steps.

\subsection{Multimodal Learning}
Originally, multimodal learning has made significant strides in computer science fields such as machine translation and computer vision \cite{multimodal_dbm, multimodal2}. Multimodal learning involves extracting attributes from multiple data streams with different shapes and dimensions, then learning to fuse these heterogeneous features and project them into a common representation space \cite{multimodal_survey}.
Two widely recognized approaches to conducting multimodal learning are early fusion and late fusion \cite{multimodal_fusion}. In late fusion, multiple decisions of classifiers are combined, while in early fusion, multiple representations from different inputs share a single hidden layer as a joint representation. In early fusion \cite{multimodal2}, feature fusion, typically achieved through layer-wise concatenation, forms a single feature map $H_3$ by horizontally combining multiple input features $H_1$ and $H_2$ as $H_3 = [H_1, H_2]$. 
In many cases where multimodal learning enhances accuracy, it does so by obtaining additional information beyond a single modality or leveraging information based on relationships between modalities. Prior study \cite{handcraftedfeatures} shows that models perform optimally when combining representations from feature extraction with human-generated features.

The success of multimodal learning in these domains has spurred its application in wider domains, such as the classification of social media activity \cite{multimodal_socialmedia, multimodal_social, multimodal_intent}, the prediction of stock prices and credit scores in finance \cite{multimodal_stockprice, alaraj}, forecasting the usage amount of smartphone games \cite{niimi_jjas}, and evaluating customer product reviews \cite{niimi_arxiv}. Many of these studies emphasize the importance of multimodal learning that considers relationships between multiple modalities. 
It should be noted that some studies employ multimodal learning using attention mechanisms and developed models such as Memory Fusion Network (MFN) \cite{memoryfusion, multimodal_attention}, while others combine mechanisms and LSTM \cite{alaraj, multimodal_AttLSTM}, STA-Transformer \cite{niimi_jjas}, and cross-attention between image and text \cite{crossAttentionDisaster}.
As mentioned, since STA and cross-attention can model relationships between the source (input) and target (output), enabling the adjustment of attention weights based on features from modalities such as tabular data by placing different modalities at the source and target. It is shown to have the better performance than feature fusion. 

\subsection{Consumer Heterogeneity and UGCs}
In marketing literature, there has long been an acknowledgement of consumer heterogeneity, defined as latent differences in behaviors among multiple consumers. These differences, stemming from unobservable attributes such as demographic variables, life stage, and purpose of visit, significantly affect observable behaviors.
The problem addressed in this study is that even when multiple users rate a restaurant similarly, understanding their preferences is hindered by the inability to discern latent differences. However, user-generated contents (UGCs), including review texts, offer potential insights into these differences.

Prior studies on electronic word-of-mouth (eWOM) \cite{wom1, wom2} and UGCs \cite{ugc1, ugc2} have predominantly focused on their impact on other consumers' brand attitudes, purchase intentions, and similar factors. However, UGCs also provide valuable information about consumer's own perceptions and attitudes toward products or services, enabling partial identification of heterogeneity without additional surveys typically required for cross-sectional data such as user profiles. 
Several studies \cite{bert_hotel, bert_googleplay, niimi_arxiv} have enhanced the accuracy of product recommendations by analyzing customer evaluation data. Nonetheless, a significant challenge for these studies is their reliance on single-modality textual data. As discussed in the Multimodal Learning subsection, extending these studies to multimodal learning models holds promise for further accuracy improvements. Moreover, one study \cite{cross-hetero} has constructed the crossmodal transfer learning model with considering heterogeneity in image and text.

\subsection{Research Gap and the Objective}
Based on previous studies, we formulate hypotheses. As mentioned above, machine learning has drastically advanced so far; however, research gap exists especially with regard to applications of machine learning in marketing.

For example, while BERT is capable of acquiring deep-contextualized word representations based on literal context, marketing context encompasses broader aspects, including consumer demographics and life stage. Regarding the consumer behaviors such as posting the review on the online platforms, the meaning of the word could depend not only on text context but also on the background information of the consumers. Surprisingly,  there are no studies addressing this broader context of textual and tabular data within a single model. Therefore, we construct a context-aware multimodal deep learning model using BERT and cross-attention, considering consumer context to predict behavior. Therefore, our first hypothesis is as follows.

\RQ{1}{The prediction accuracy improves significantly using the context-aware model compared to reference models.}

Additionally, we assess the effectiveness of our model across diverse sample groups: Restaurants, Nightlife, and \cafe ~(cf. Data Description subsection). Given the diverse characteristics of the Nightlife category, which may include entertainment factors such as shows, music, and alcohol, predicting ratings in this category is expected to be more challenging. Thus, we propose the following hypothesis.

\RQ{2}{The prediction accuracy decreases on average in Nightlife category.}

Moreover, multimodal learning models often contend with numerous parameters and complex architectures. In such sparse training scenarios, determining which parameters to update can be challenging, leading to vanishing gradient problem. While Adam optimization is a common choice for deep learning, Adamax may prove more effective in training such models since the original paper \cite{adam} highlights out the advantage in sparse gradients. Therefore, we establish another hypothesis as follows.

\RQ{3}{In comparing the performance of multiple optimizers, Adamax achieves the highest average test score.}

Although prior studies have compared different forms of LLMs in terms of prediction accuracy, the research landscape in applied domains remains somewhat lacking. In particular, BERT comes in various forms, distinguished by scale ($params_{BERT}$: number of parameters in BERT) and advanced models (e.g., RoBERTa and DistilBERT). The question arises regarding BERT's impact on performance: whether it merely serves as a means of acquiring deep-contextualized word representations, or if prediction accuracy can be further improved by employing more advanced or larger-scale BERT models. Therefore, we establish two hypotheses, respectively.

\RQ{4-1}{The prediction accuracy improves on average with larger-scale pre-training models.}

\RQ{4-2}{The prediction accuracy improves on average with newer pre-training models.}

\begin{figure}[ht]
\centering
   \includegraphics[width=0.9\linewidth]{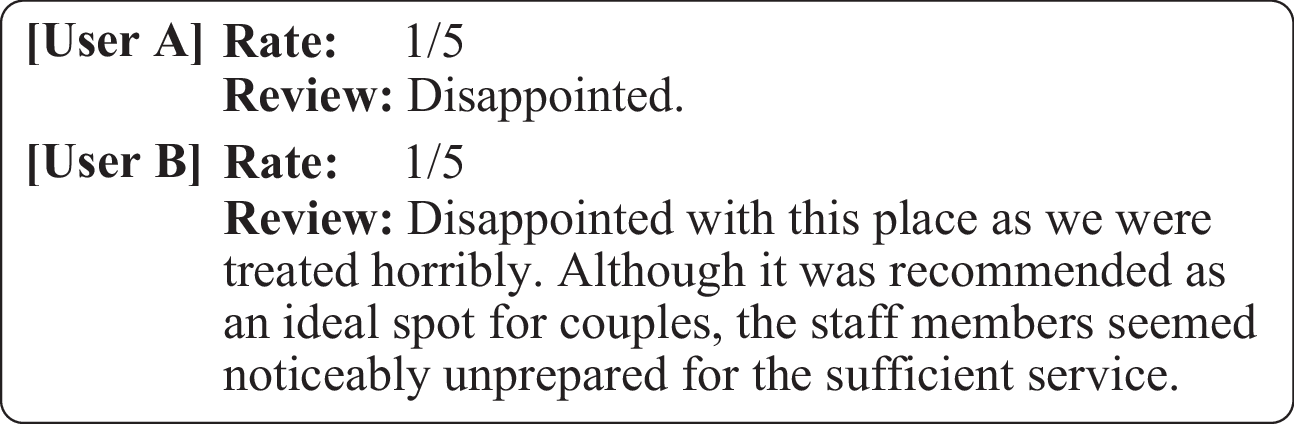}
      \caption{Amount of information in the review text}
      \label{fig:difference}
\end{figure}

Lastly, unlike tabular data, textual data differs significantly in the amount of information conveyed in each post (e.g., while one post contains only one-word impressions, another might provide detailed information about the user's situation and background). This variability poses a challenge for prediction accuracy, as illustrated by the differing amounts of information that the post retains.
For example, regarding the two review texts shown in Fig. \ref{fig:difference}\footnote{Since the dataset employed in this study prohibits the disclosure of actual review sentences, the texts shown in the figure are fictitious ones written by the authors.}, even though they are all text data, the amount of information each holds is completely different. 
Similar issues have been pointed out in marketing fields, for example, one study \cite{rfm-clv} pointed out that the prediction using behavioral logs may vary in accuracy depending on the quantity of services used. 
This problem is anticipated to arise in multimodal learning with textual data as well. Therefore, in this study, we also assess the change in prediction accuracy based on the number of tokens in the textual input. Thus, we propose the following hypothesis.

\RQ{5}{Prediction accuracy decreases with fewer tokens in multimodal learning.}

\section{Model}
\subsection{Architecture}
\begin{figure}[ht]
      \centering
      \includegraphics[keepaspectratio,width=0.8\linewidth]{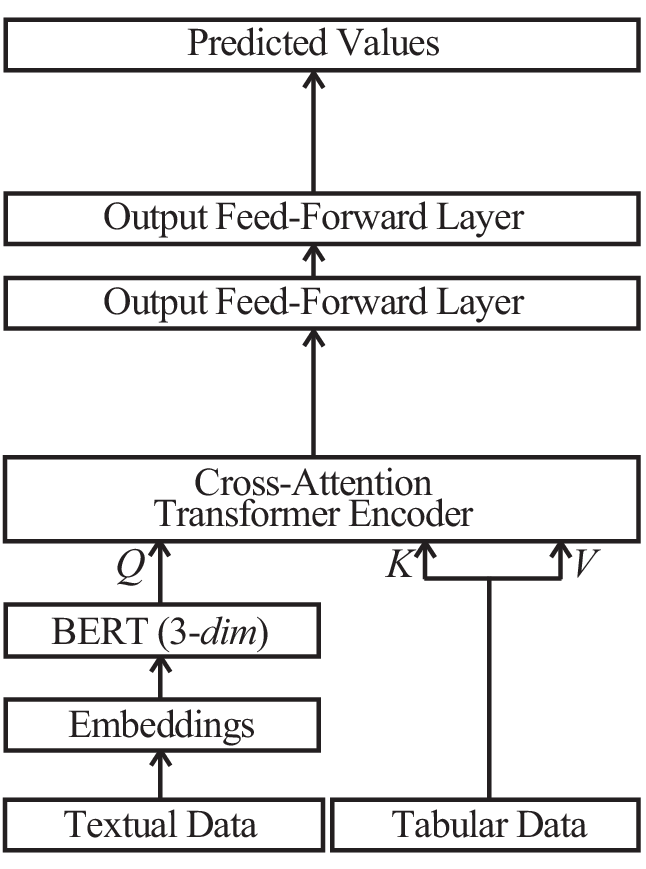}
      \caption{Context-Aware Model}
      \label{fig:architecture1}
\end{figure}

This study addresses both textual and tabular data which needs multiple inputs. The network is divided into three subnets based on their roles: X1-, X2-, and Output-subnet. X1- and X2-subnet process each modality with appropriate structures, and Output-subnet concatenates them and predicts the values in the upper layers (Fig. \ref{fig:architecture1}; $bs$: batch size, $J$: number of tabular variables).

First, in the X1-subnet handling textual data, BERT and a tokenizer are employed to acquire deep-contextualized word representations. As discussed, using the pooler-output in multimodal learning may not always be optimal as it could lead to dimensionality reduction based on the [CLS] token in BERT. Despite the possible for selecting necessary features through multihead attentions within cross-attention, in this model, we opt for the state of the final hidden layer in BERT as the BERT output with dimensions ($bs$, $len_{max}$, $param_{\BERT}$).
Second, in the X2-subnet managing tabular data, while feed-forward layers can be incorporated, the input data should not be overly processed prior to feature fusion. Therefore, we choose to directly feed the input into the Output-subnet.

The Output-subnet receives these two representations, fuses them, and generates outputs. While prior studies \cite{alaraj, niimi_jjas} have utilized both STA and layer-wise concatenation, if the STA mechanism adequately captures the features of two modalities, it is uncertain whether early fusion is necessary to obtain a joint representation. Hence, our proposed model adopts the cross-attention Transformer encoder with eight attention heads. This mechanism is anticipated to yield high accuracy without feature fusion, as it captures the relationship between the two modalities from multiple perspectives, which can be challenging with a single attention mechanism. In this study, this proposed model is called the context-aware model.

\subsection{Evaluation}
\begin{figure}[t]
   \centering
\subfloat[][Context-fusion model]{\includegraphics[width=0.48\linewidth]{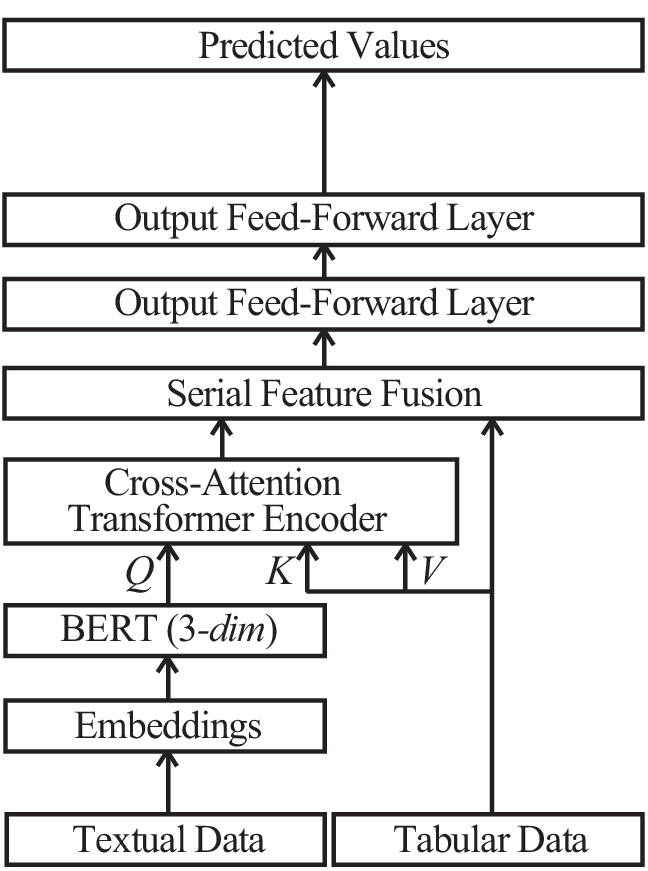}\label{fig:architecture2}} 
\quad
\subfloat[][Feature-fusion model]{\includegraphics[width=0.48\linewidth]{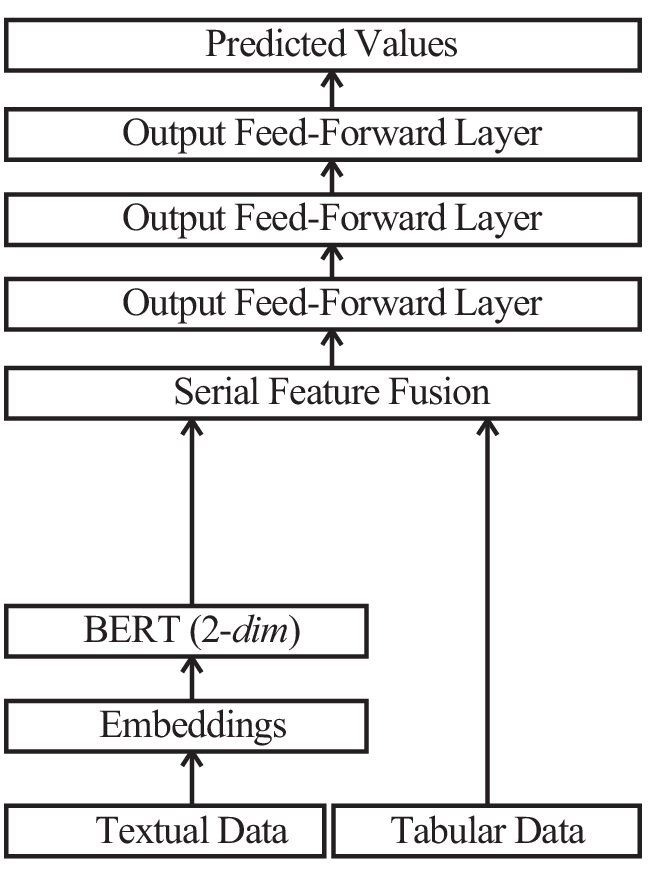}\label{fig:architecture3}} 
   \caption{Reference models (multimodal)}
   \label{fig:architecture_multimodal}
\end{figure}

To evaluate the effectiveness of our proposed model, we construct two multimodal learning models and two monomodal models as reference points. First, for the multimodal approach, we introduce a context-fusion model (referred to Fig. \ref{fig:architecture2}), which integrates feature fusion into the context-aware model. Additionally, we implement a typical multimodal model with feature fusion using pooler-output (refer to Fig. \ref{fig:architecture3}) \footnote{We employ pooler-output for the feature-fusion model due to the requirement of two-dimensional representations for layer-wise concatenation}.Since the feature-fusion model directly receives the high-dimensional representation from the BERT output, the number of hidden layers in the output layer post feature fusion is increased to three, with each layer comprising 512, 256, and 128 units.

\begin{figure}[t]
   \centering
\subfloat[][Textual model]{\includegraphics[width=0.48\linewidth]{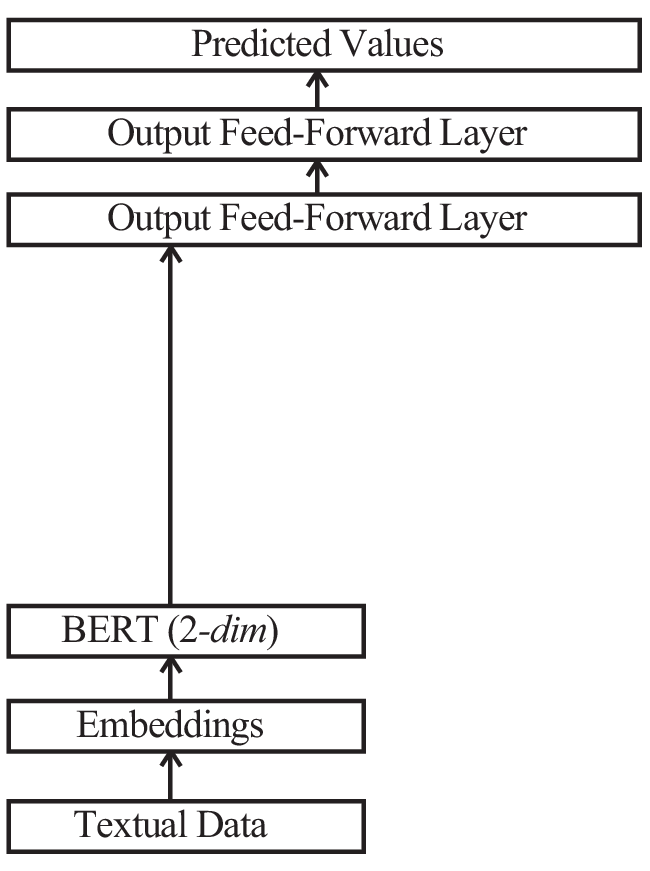}\label{fig:architecture4}} 
\quad
\subfloat[][Tabular model]{\includegraphics[width=0.48\linewidth]{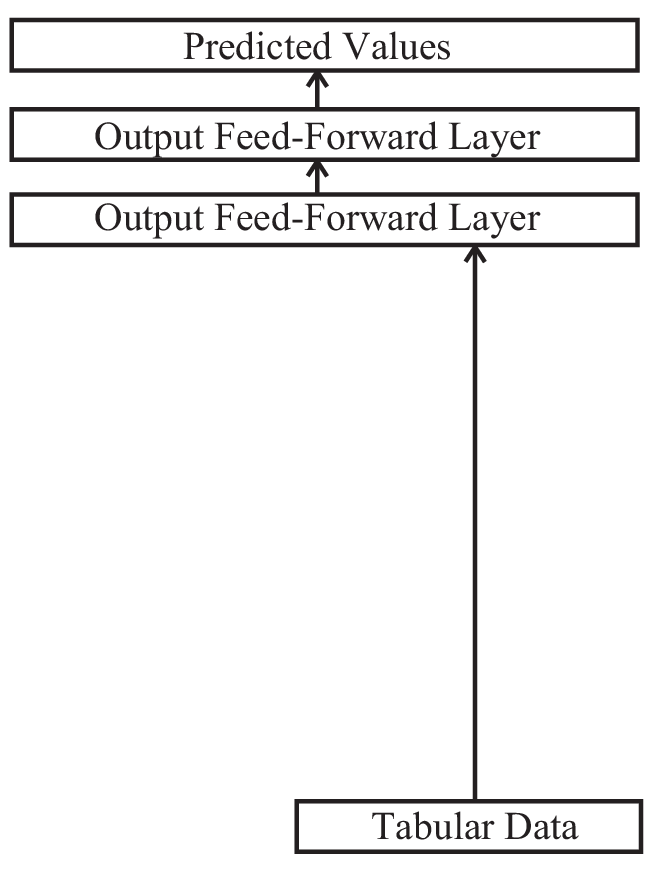}\label{fig:architecture5}} 
   \caption{Reference models (monomodal)}
   \label{fig:architecture_monomodal}
\end{figure}

Second, for the monomodal models, we introduce the textual and  tabular models (referred to Fig. \ref{fig:architecture4}, \ref{fig:architecture5}), which process modality-specific layers and transmit them to the output layer without traversing through Transformer or feature-fusion architectures.
Moreover, we incorporate two benchmarks: a linear regression model that captures linear relationships and a random model that generates random predictions within the range of $[0,1]$, and. These six reference models allows for a comprehensive comparison of the performance of the proposed models.

In terms of optimization, many existing studies have adopted Adam \cite{adam} as an optimizer; however, as described in {\bf H2}, it has yet to be clarified what optimizer is effective for a complex architecture of multimodal learning. Hence, this study delves into the impact of different optimizers on prediction accuracy, including Adam, Nesterov-accelerated Adaptive Moment Estimation (Nadam) \cite{nadam}, and Adamax. 

Furthermore, regarding the pre-trained BERT model, we initially employ bert-base-uncased among several pre-trained models to demonstrate the superior prediction accuracy of our proposed model's architecture compared to others (Study 1). Subsequently, we explore changes in accuracy by replacing bert-base-uncased with different pre-trained models (Study 2).

\begin{table}[ht] 
      \begin{center}
      \caption{Model Settings}
      \label{tab:explore}
        \scalebox{0.845}{
        \begin{threeparttable}
\begin{tabular}{llllll}
\toprule
\multicolumn{3}{c}{Parameters}        &   \multicolumn{1}{c}{Values}  \\
\midrule
\multicolumn{4}{l}{\bf Hyper-Parameters} \\
\cmidrule[0pt](l){1-2}
& \multicolumn{2}{l}{Number of Epochs} & 500\\
& \multicolumn{2}{l}{Batchsize} &         256 \\
& \multicolumn{2}{l}{Optimizer} &  Adamax \\
& \multicolumn{2}{l}{Loss Function} &  mean squared error (MSE) \\
\midrule
\multicolumn{4}{l}{{\bf X1-subnet}} \\
\cmidrule[0pt](l){1-2}
& \multicolumn{2}{l}{Structure} & BERT \\ 
& \multicolumn{2}{l}{Pre-trained Model} & bert-base-uncased\\
& \multicolumn{2}{l}{$params_{\BERT}$} & 768 \\
\midrule
\multicolumn{4}{l}{{\bf X2-subnet}} \\
\cmidrule[0pt](l){1-2}
& \multicolumn{2}{l}{Number of Hidden Layers} & 0 ~(i.e., directly connected to the upper layer.) \\
& \multicolumn{2}{l}{Number of Input Features ($J$)} & 15 \\
\midrule
\multicolumn{4}{l}{{\bf Feature Fusion}} \\
\cmidrule[0pt](l){1-2}
& \multicolumn{2}{l}{Cross-Attention} & used in 'Context-Aware' and 'Context-Fusion' \\
&& - Attention Heads &  8\\
& \multicolumn{2}{l}{Layer-wise Concatenation} & used in 'Context-Fusion' and 'Feature Fusion' \\
\midrule
\multicolumn{4}{l}{{\bf Output-subnet}} \\
\cmidrule[0pt](l){1-2}
& \multicolumn{2}{l}{Activation (hidden layers)} & $tanh$ \\
& \multicolumn{2}{l}{Activation (output)} & $tanh$ \\
& \multicolumn{2}{l}{Hidden Layers} & 2-3 (cf. Fig. \ref{fig:architecture}) \\
\bottomrule
\end{tabular}
\begin{tablenotes}[para,flushleft,online,normal] 
{\it Note.} $tanh$ stands for hyperbolic tangent function.
\end{tablenotes}
\end{threeparttable}
        }
        \end{center}
\end{table} 

\subsection{Data Description}
To validate the efficacy of the proposed model, we require behavioral log data containing both textual and tabular information. For this purpose, we utilize the Yelp Open Dataset \cite{yelp}. Yelp, an online platform, offers a wealth of information about various venues including restaurants, stores, and public facilities, alongside user ratings and reviews. The dataset comprises user review texts, profiles, and venue details. Each location is associated with one or more category tags, facilitating the extraction of target locations by specifying these tags. In our study, we focus on three business categories to demonstrate the robustness of the model: Restaurants (tagged with ``Restaurants'', but not neither with ``Fast Food'', ``Food Truck'', ``Nightlife'', and ``Bar''), Nightlife (tagged with both ``Restaurants'' and ``Nightlife'', but not neither with ``Fast Food'' and ``Food Truck''), and \cafe~(tagged with both ``Cafes'' and ``Coffee and Tea'', but not neither with ``Fast Food'' and ``Food Truck''). In particular, Nightlife category encompasses various types of establishments such as bars and nightclubs, posing challenges in evaluation solely based on store information. 

For the sake of data acquisition convenience, we predict the ratings (i.e., the number of stars) of restaurants using review texts, user profile information, and restaurant information. While the target variable can be readily obtained from the app, its accurate prediction by our proposed model signifies its suitability in understanding consumer preferences and its potential extension into an effective recommender system \cite{bert_hotel}. 

We randomly sample 10,000 posts of ratings and reviews from each category containing one or more English words in year 2018. In cases where a user reviews the same location multiple times, we consider only the latest post. For textual data preprocessing, we replace line breaks, emojis, icons, and other symbols with periods and merge continuous sequences of periods into a single period. A summary of the dataset statistics is provided in Table \ref{tab:stats}. Notably, there are no duplications for the location in the Restaurants and Nightlife categories.

\begin{table}[ht]
      \begin{center}
      \caption{Statistics of the Categories}
      \label{tab:stats}
        \scalebox{1}{
        \begin{threeparttable}
\begin{tabular}{p{1.1cm}
wc{0.5 cm} wc{0.6 cm} wc{0.5 cm} wc{0.3 cm}
wc{0.5 cm} wc{0.4 cm} wr{0.4 cm} wc{0.4 cm}}
\toprule
\multirow{2}{*}{Category} & \multirow{2}{*}{\#Users} & \multirow{2}{*}{\#Spots} &  \ccol{2}{\#Stars} & \ccol{4}{\#Tokens} \\
\cmidrule(lr){4-5} \cmidrule(l){6-9}
&  & & Mean & Std & Mean & Std & Min & Max \\
\midrule
Restaurants&9567 & 1387   & 3.9 &  1.4 & 110.3 &  ~~91.6 &   9 &       512 \\
Nightlife     & 9491 & 2097   & 3.9 &  1.4 & 119.9 &   100.1 &  11 &      512 \\
\cafe          & 9189 & ~~665 & 4.2 &  1.2 & 109.9 & ~~90.5 & 10 &      512 \\
\bottomrule
\end{tabular}
\begin{tablenotes}[para,flushleft,online,normal] 
{\it Note.} \#Users and \#Spots indicate the unique numbers of users and restaurants in each category, respectively.
\end{tablenotes}
\end{threeparttable}
        }
        \end{center}
\end{table} 

The dataset of each category consists of $D = \{(x_i, y_i)\}_{i=1}^{n}$ with a sample size of $n=10000$, where each input $x_i$ comprises one textual variable and $J$ tabular variables, denoted as $x_i = (x_i^{(text)}, x_i^{(tab)}) = (x_{1i}^{(text)}, x_{1i}^{(tab)}, x_{2i}^{(tab)}, \dots, x_{Ji}^{(tab)}) \in \mathbb{X}$. The target variable $y_i \in \mathbb{Y} = [0,1]$ represents a normalized value of the ratings, which is scaled between 0 and 1 from a range of 1 to 5 stars. The variables are shown in Table \ref{tab:varlist}.

The dataset of 10,000 observations is split into training (70\%), validation (15\%), and test (15\%) subsets. During training, the loss function employed is mean squared error (MSE), while model performance is evaluated using root mean squared error (RMSE) using actual and predicted values $(y_i, \hat{y}_i)$ as follows:
\begin{align}
  \text{MSE} &= \frac{1}{n}\sum_{i=1}^{n} (y_i - \hat{y}_i)^2\\
  \text{RMSE} &= \sqrt{\text{MSE}}
\end{align}
Detailed model settings are provided in Table \ref{tab:explore}.

\begin{table}[t] 
\small
      \begin{center}
      \caption{Model Variables}
      \label{tab:varlist}
        \scalebox{1}{
        \begin{threeparttable}
\begin{tabular}{
p{1.8cm}
p{5.6cm}
}
\toprule
\ccol{1}{Variable Name} & \multicolumn{1}{c}{Description} \\
\midrule
\multicolumn{2}{l}{$\bf Y$~~: Target Variable ($bs$, $1$)} \\
\cmidrule[0pt](l){1-2}
- $rating$ &  Rating value posted on Yelp $^\dagger$ \\
\midrule
\multicolumn{2}{l}{$\bf \Xone$: Textual Variable ($bs$, $1$)} \\
\cmidrule[0pt](l){1-2}
- $review$ &  Review text posted on Yelp tokenized\\
                     & with fixed-length of $len_{max}=512$ tokens \\
\midrule
\multicolumn{2}{l}{$\bf \Xtwo$: Tabular Variables shape ($bs$, $15$)} \\
\cmidrule[0pt](l){1-2}
\multicolumn{2}{l}{Location} \\
- $open\_dow$ & A number of days of week opening $^\dagger$\\
- $open\_hours$ & A number of opening hours in a week $^\dagger$\\
- $open\_mon$ & A number of opening hours in Monday $^\dagger$\\
~~\dots & \dots\\
- $open\_sun$ & A number of opening hours in Sunday $^\dagger$\\
\cmidrule[0.1pt](l){1-2}
\multicolumn{2}{l}{User} \\
- $n\_f\hspace{-0.1em}riends$ & A number of friends \\
- $n\_f\hspace{-0.1em}ans$ & A number of getting fans\\
- $n\_elites$ & A number of getting elite  \\
\cmidrule[0.1pt](rl){1-2}
\multicolumn{2}{l}{Post} \\
- $n\_use\hspace{-0.1em}f\hspace{-0.1em}ul$ & A number of getting useful \\
- $n\_f\hspace{-0.1em}unny$ & A number of getting funny \\
- $n\_cool$ & A number of getting cool  \\
\bottomrule
\end{tabular}
\begin{tablenotes}[para,flushleft,online,normal] 
$^\dagger$Variables are normalized in $[0,1]$.
\end{tablenotes}
\end{threeparttable}
        }
        \end{center}
\end{table} 

\section{Results and Discussion}
\subsection{Study 1: Comparison Across the Model Architectures}\label{sec:study1}
The results are presented in Table \ref{tab:result1}, reveal a similar pattern across all categories. The proposed model consistently achieves the highest prediction accuracy in the test scores across all categories. The context-fusion model follows closely behind, while the performance of the feature-fusion model sometimes lags behind that of the textual and even linear regression models.
In particular, despite the context-fusion model having the largest number of parameters in this study. Context-Fusion model fuses the representations twice with STA-Transformer and feature fusion, but the contribution on the performance is actually limited. Conversely, the random model exhibits the lowest accuracy, followed by the linear regression model in most cases.

\begin{figure}[ht]
   \centering
   \includegraphics[width=0.98\linewidth]{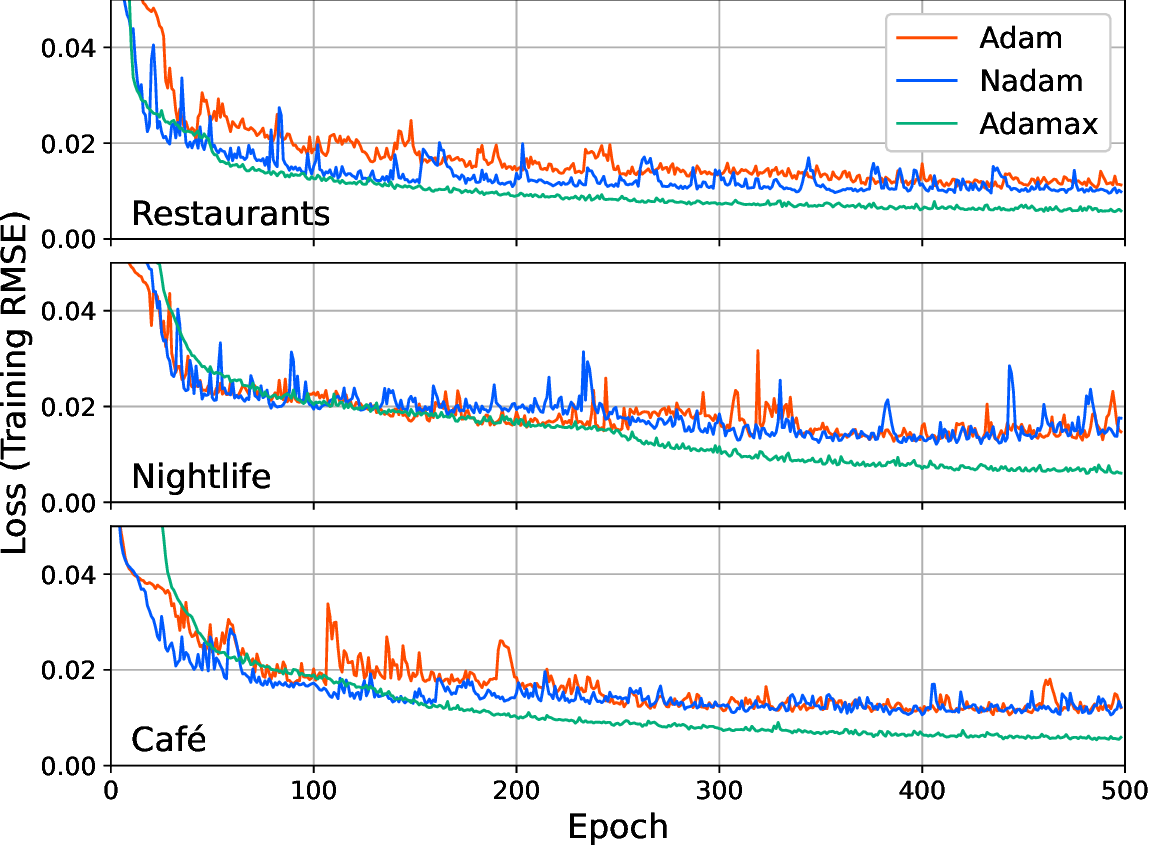}
   \caption{Training process by different optimizers}
   \label{fig:graph_optimizer}
\end{figure}

Although some reference models stopped training in fewer epochs than the proposed model, this trend does not necessarily indicate early convergence due to the absence of early stopping \cite{earlyStopping}. Rather, it suggests that these models struggled to escape local convergence in the early stages \footnote{A similar tendency is confirmed in Fig. \ref{fig:graph_optimizer}. Noted that it is about a different analysis.}. These results guarantee the generalized performance of the proposed model, and thus, {\bf H1} is supported. 

\begin{table*}[t] 
      \begin{center}
      \caption{Results (with Adamax optimizer, ascending in Test RMSE)}
      \label{tab:result1}
      \scalebox{0.98}{
        \begin{threeparttable}
\begin{tabular}{
p{0.1cm}
p{1.45cm}
p{2.1cm}
wc{1.9cm}
wc{1.4cm}
wc{0.4cm}
wc{1.3cm}
wc{0.5cm}
wc{0.8cm}
wc{1.8cm}
wr{1.7cm}
}
\toprule &
    \ccol{1}{Model} & \ccol{1}{Modality} & BERT Model & Optimizer & 
    Train & Validation & Test & 
    Epochs & Training Time & \ccol{1}{\#Parameters} \\
\midrule
\parbox[t]{2mm}{\multirow{7}{*}{\rotatebox[origin=c]{90}{\bf Restaurant}}}
& Multimodal & context-aware &  bert-base-uncased & adamax & \bf0.085 & 0.135 & \bf0.132 & 316 & 04:01:19 & 119,122,520 \\
& Multimodal & context-fusion &  bert-base-uncased & adamax & 0.109 & \bf0.130 & 0.134 & \bf125 & 01:50:58 &  119,123,080 \\
& X1-modal & textual               & bert-base-uncased & adamax & 0.151 & 0.149 & 0.143 & 499 & 05:53:48 & 109,712,129 \\
& Multimodal & feature-fusion  &  bert-base-uncased & adamax & 0.152 & 0.152 & 0.155 & 290 & 03:33:04 & 110,048,001 \\
& X2-modal & tabular               & bert-base-uncased & adamax & 0.258 & 0.260 & 0.261 & 313 & \bf00:01:05 & 281 \\
\cmidrule[0.1pt]{4-8}
& & & \multicolumn{2}{r}{Linear regression:} & 0.259 &  0.261 &  0.262 \\
& & & \multicolumn{2}{r}{Random:} & 0.494 &  0.496 &  0.503 \\
\midrule
\parbox[t]{2mm}{\multirow{7}{*}{\rotatebox[origin=c]{90}{\bf Nightlife}}}
& Multimodal & context-aware &  bert-base-uncased & adamax & \bf0.084 & \bf0.127 & \bf0.140 & \bf401 &  05:22:41 & 119,122,520 \\
& Multimodal & context-fusion &  bert-base-uncased & adamax & 0.093 & 0.129 & 0.141 & 406 & 05:25:48 & 119,123,080 \\
& X1-modal & textual              & bert-base-uncased & adamax & 0.150 & 0.141 & 0.150 & 476 & 05:33:54 & 109,712,129 \\
& Multimodal & feature-fusion  &  bert-base-uncased & adamax & 0.141 & 0.144 & 0.161 & 423 & 05:02:02 & 110,048,001 \\
& X2-modal & tabular                & bert-base-uncased & adamax & 0.255 & 0.254 & 0.257 & 471 & \bf00:01:31 & 281 \\
\cmidrule[0.1pt]{4-8}
& & & \multicolumn{2}{r}{Linear regression:} & 0.262 &  0.259 &  0.260 \\
& & & \multicolumn{2}{r}{Random:} & 0.481 &  0.480 &  0.482 \\
\midrule
\parbox[t]{2mm}{\multirow{7}{*}{\rotatebox[origin=c]{90}{\bf \cafe}}}
& Multimodal & context-aware &  bert-base-uncased & adamax & 0.076 &  0.127 & \bf0.125 & 475 &  06:08:27 & 119,122,520 \\
& Multimodal & context-fusion &  bert-base-uncased & adamax & \bf0.074 & \bf0.125 &  0.127 & 480 &  06:02:39 & 119,123,080 \\
& Multimodal & feature-fusion  &  bert-base-uncased & adamax & 0.137 &  0.147 &  0.142 & \bf213 &  02:43:10 & 110,048,001 \\
& X1-modal & textual                & bert-base-uncased & adamax &  0.137 &  0.141 &  0.147 &    491 &  06:10:34 &  109,712,129\\
& X2-modal & tabular                & bert-base-uncased & adamax & 0.231 &  0.241 &  0.228 &    234 & \bf00:01:03 &     281 \\
\cmidrule[0.1pt]{4-8}
& & & \multicolumn{2}{r}{Linear regression:} & 0.231 &  0.236 &  0.223 \\
& & & \multicolumn{2}{r}{Random:} & 0.509 &  0.500 &  0.516 \\
\bottomrule
\end{tabular} 
\begin{tablenotes}[para,flushleft,online,normal] 
{\it Note.} Bold type represents the best model for the indices. Training Time shows the actual duration for the best validation score in $hh$:$mm$:$ss$ using the same environment (GPU: NVIDIA A100-SXM4-40GB).
\end{tablenotes}
\end{threeparttable}
}
        \end{center}
\end{table*} 

Second, Table \ref{tab:average_result} provides an overview of the average performance considering various perspectives: such as target categories, modalities, and optimizers. As anticipated, the Nightlife category exhibits slightly lower test performance than the Restaurants category, possibly due to the diverse nature of establishments in the Nightlife category. Nonetheless, the mean score by modality indicates a high level of predictability, underscoring the usefulness of analyzing multiple modalities. This result supports {\bf H2}.

\begin{table}[t] 
      \begin{center}
      \caption{Average Performances by the Group}
      \label{tab:average_result}
      \scalebox{0.9}{
        \begin{threeparttable}
\begin{tabular}{
wl{1.6cm}
wc{0.3cm}
wc{0.9cm}
wc{0.3cm}
wc{0.7cm}
wc{1.4cm}
wl{1.4cm}
}
\toprule
 
    \ccol{1}{Group} & Train & Validation & Test & 
    Epochs & Training Time & \ccol{1}{\#Parameters}  \\
\midrule
\multicolumn{3}{l}{\bf Category} \\
\cafe      &  \bf0.159 & \bf0.178 &  \bf0.175 &  340.5 &  03:14:07 &  91,601,202.2 \\
Restaurants    &  0.193 &  0.207 &  0.209 &\bf278.5 &  \bf02:48:33 &  91,601,202.2 \\
Nightlife      &  0.208 &  0.218 &  0.226 &  321.4 &  03:07:13 &  91,601,202.2 \\
\midrule
\multicolumn{3}{l}{\bf Modality} \\
context-aware & \bf0.100 & \bf0.133 & \bf0.140 &  394.7 &  05:01:51 &  119,122,520.0 \\
context-fusion & 0.108 &  0.133 &  0.141 &  400.2 &  05:02:31 &  119,123,080.0 \\
Multimodal &  0.153 &  0.174 &  0.179 &  320.9 &  04:02:31 &  116,097,867.0 \\
X1-modal   &  0.228 &  0.230 &  0.229 &  312.7 &  03:08:03 &  109,712,129.0 \\
X2-modal   &  0.249 &  0.253 &  0.251 &  292.0 & \bf00:00:53 &        281.0 \\
feature-fusion &  0.250 &  0.255 &  0.257 & \bf167.8 &  02:03:12 &  110,048,001.0 \\
\midrule
\multicolumn{3}{l}{\bf Optimizer} \\
Adamax    & \bf0.142 & \bf0.160 & \bf0.163 &  374.2 &  03:29:54 &  91,601,202.2 \\
Nadam      & 0.202 &  0.214 &  0.217 &  309.1 &  \bf02:48:02 &  91,601,202.2 \\
Adam        & 0.217 &  0.228 &  0.231 &  \bf257.1 &  02:51:57 &  91,601,202.2 \\
\bottomrule
\end{tabular} 
\begin{tablenotes}[para,flushleft,online,normal] 
{\it Note.} Bold type represents the best model for the indices. Training Time shows the actual duration for the best validation score in $hh$:$mm$:$ss$ using the same environment (GPU: NVIDIA A100-SXM4-40GB).
\end{tablenotes}
\end{threeparttable}
}
        \end{center}
\end{table} 

Third, the results in Table \ref{tab:average_result} further demonstrate the effectiveness of adamax as an optimizer, particularly in handling the complex structure of neural networks dealing with sparse textual representations. Despite taking longer for training, adamax proves considerably more effective. Notably, even with an enormous number of parameters, adamax demonstrates superior performance in effectively updating the weights.
A comparison of the change in losses of the context-aware model among different optimizers in Fig. \ref{fig:graph_optimizer}, corroborates these findings. Adamax shows outstanding effectiveness over the training epochs. The progression of learning in the three categories shows that adamax does not always train efficiently from the early stages; however, as it proceeds to the later stages, only adamax continues to reduce the loss while the other optimizers converges locally. Thus, {\bf H3} is supported.

\subsection{Study 2: Impact of Replacing Pre-Trained Models}\label{sec:study2}
The results from Study 1 demonstrate the effectiveness of our proposed architecture; however, even with its high accuracy in multimodal learning, the model relies on BERT-Base-Uncased component. To further investigate the impact of different pre-trained models, we conducted additional analyses by replacing the BERT component with BERT-Large-Uncased, RoBERTa-Base, and RoBERTa-Large within context-aware model.

The findings, presented in Table \ref{tab:result_bert}, confirm a significant improvement in test performance on average with BERT-Large-Uncased and RoBERTa-Base compared to Bert-Base-Uncased across all three categories. Both RoBERTa-Base and BERT-Large-Uncased contribute to the accuracy, while RoBERTa-Large does not exhibit the same level of improvement. The average test scores suggest that both RoBERTa-Base and BERT-Large-Uncased demonstrate comparable generalization capabilities, with RoBERTa-Base outperforming in terms of convergence time.
The lower accuracy observed with the RoBERTa-Large component could be attributed to the insufficient sample size relative to the complexity of the architecture. Previous studies have indicated that large-scale models like RoBERTa-Large require a larger sample size for optimal performance. Thus, both {\bf H4-1} and {\bf H4-2} are supported, respectively.

\begin{table*}[ht] 
      \begin{center}
      \caption{Impact of the pre-trained models (with Adamax optimizer, ascending in Test RMSE)}
      \label{tab:result_bert}
      \scalebox{1.1}{
        \begin{threeparttable}
\begin{tabular}{
p{0.2cm}
p{1.6cm}
p{2.3cm}
wc{0.7cm}
wc{1cm}
wc{1cm}
wc{1cm}
wc{0.8cm}
wc{1.6cm}
wc{1.5cm}
}
\toprule &
    \ccol{1}{Modality} & \ccol{1}{BERT Model} & Optimizer & 
    Train & Validation & Test &
    Epochs & Training Time & \#Parameters \\
\midrule
\parbox[t]{2mm}{\multirow{4}{*}{\rotatebox[origin=c]{90}{Restaurant}}}
& context-aware &  bert-large-uncased & adamax & 0.079 &  \bf0.129 &  \bf0.121 &    496 &  16:16:07 &  347,927,896 \\
& context-aware &  roberta-base & adamax & 0.091 & 0.133 & 0.131 & 319 & 04:11:54 & 134,285,912 \\
& context-aware &  bert-base-uncased & adamax & 0.085 & 0.135 & 0.132 & \bf316 & \bf04:01:19 & 119,122,520 \\
& context-aware &  roberta-large & adamax & \bf0.076 &  0.144 &  0.150 &    490 &  16:00:06 &  368,145,752 \\
\midrule
\parbox[t]{2mm}{\multirow{4}{*}{\rotatebox[origin=c]{90}{Nightlife}}}
& context-aware &  roberta-base & adamax & 0.095 & \bf0.120 & \bf0.130 & \bf176 & \bf02:23:58 & 134,285,912 \\
& context-aware &  bert-large-uncased & adamax & 0.085 & 0.123 & 0.135 & 416 & 16:29:20 & 347,927,896 \\
& context-aware &  roberta-large & adamax & \bf0.084 & 0.125 & 0.136 & 401 & 13:14:30 & 368,145,752 \\
& context-aware &  bert-base-uncased & adamax & 0.084 & 0.127 & 0.140 & 401 & 05:22:41 & 119,122,520\\
\midrule
\parbox[t]{2mm}{\multirow{4}{*}{\rotatebox[origin=c]{90}{\cafe}}}
& context-aware &  roberta-base & adamax & 0.077 & \bf0.118 & \bf0.120 & 377 & \bf04:51:41 &  134,285,912 \\
& context-aware &  bert-large-uncased & adamax & \bf0.071 &  0.122 &  0.124 &    494 &  16:16:29 &  347,927,896 \\
& context-aware &  bert-base-uncased & adamax & 0.076 &  0.127 &  0.125 &    475 &  06:08:27 & 119,122,520 \\
& context-aware &  roberta-large & adamax & 0.082 &  0.128 &  0.137 & \bf238 &  07:58:23 &  368,145,752 \\
\midrule
\parbox[t]{2mm}{\multirow{4}{*}{\rotatebox[origin=c]{90}{Average}}}
& context-aware &  bert-large-uncased & adamax & \bf0.078 &  0.125 &  \bf0.127 &  468.7 &  16:20:39 &  347,927,896 \\
& context-aware &  roberta-base & adamax & 0.088 &  \bf 0.123 &  0.127 &  \bf290.7 &  \bf03:49:11 &  134,285,912 \\
& context-aware &  bert-base-uncased & adamax & 0.082 &  0.130 &  0.132 &  397.3 &  05:10:49 &  119,122,520 \\
& context-aware &  roberta-large & adamax & 0.081 &  0.132 &  0.141 &  376.3 &  12:24:20 &  368,145,752 \\
\bottomrule
\end{tabular} 
\begin{tablenotes}[para,flushleft,online,normal] 
{\it Note.} Bold type represents the best model for the indices. Training Time shows the actual duration for the best validation score in $hh$:$mm$:$ss$ using the same environment (GPU: NVIDIA A100-SXM4-40GB).
\end{tablenotes}
\end{threeparttable}
}
        \end{center}
\end{table*} 

\subsection{Study 3: Impact of the Number of Tokens}\label{sec:study3}
Finally, we examine the impact of the amount of information in the review text on prediction accuracy, as described in {\bf H4-1} and {\bf H4-2}. We regard the number of tokens in the review as a measure of information and investigate whether accuracy varies with the number of tokens. The best model from Study 1 (context-aware model with bert-base-uncased and the adamax optimizer) is utilized for each category. We set up the token strata by dividing three subsets of training, validation, and test data into 20\% according to the number of tokens. Then, we predict and compute the average RMSE by strata.

The results, categorized by the number of tokens and by stratum are shown in Table \ref{tab:result_token}\footnote{Note that the values shown in the table represent the prediction accuracy for trained, validated, and test samples without any further training.}. For the test data alone, the prediction accuracy is highest when the number of tokens is lowest in the Restaurants category, and medium in the other two categories. This suggests that while multimodal learning of textual and tabular data is expected to improve prediction accuracy, it does not always require a large amount of text-based information. For two categories other than Restaurants, the prediction accuracy is also best in the lowest tokens strata in training and validation. However, in the Nightlife and \cafe categories, which exhibit wide variation in location attributes, higher numbers of tokens ensure generalizability in test performance, whereas the model for the Restaurants category demonstrates high generalizability with fewer tokens.
In addition, regarding the observed decrease in accuracy with particularly large numbers of tokens, several possible reasons exist. First, we cut off sentences with more than 512 tokens due to the size of BERT's context window, which may not convey enough information to the model. Second, excessively long texts may contain redundant information unrelated to the user ratings, leading to that the model has not properly discerned the information. Thus, {\bf H5} is not supported.

\begin{table}[ht]
      \begin{center}
      \caption{Impact of the Number of Tokens (with Adamax optimizer, ascending in the Number of Tokens)}
      \label{tab:result_token}
        \begin{threeparttable}
      \scalebox{0.94}{
\begin{tabular}{crccrccrcc}
\midrule
& \ccol{3}{Train} & \ccol{3}{Validation} & \ccol{3}{Test} \\
\cmidrule(r){2-4} \cmidrule(lr){5-7} \cmidrule(l){8-10} 
& \ccol{2}{$M$} & RMSE & \ccol{2}{$M$} & RMSE & \ccol{2}{$M$} & RMSE  \\
\midrule
\parbox[t]{2mm}{\multirow{5}{*}{\rotatebox[origin=c]{90}{Restaurants}}}
 & ~~~~31.5 &&  \bf0.067 & ~~~~30.4 && \bf0.109 & ~~~~31.8 &&   \bf0.115  \\
&  54.1 &&        0.068 & ~~54.3 &&      0.130 &  ~~55.5 &&       0.125 \\
&  81.8 &&        0.069 & ~~81.8 &&      0.130 & ~~82.9 &&       0.127 \\
&     127.1 &&        0.075 &   125.7 &&      0.135 &    131.8 &&       0.147 \\
&     258.2 &&        0.086 &   262.7 &&      0.165 &    260.3 &&       0.154 \\
\midrule
\parbox[t]{2mm}{\multirow{5}{*}{\rotatebox[origin=c]{90}{Nightlife}}}
& 31.8 &&     \bf0.125 & ~~33.3 &&  \bf0.118 &  ~~33.0 &&       0.137 \\
& 56.2 &&        0.132 & ~~56.7 &&      0.130 & ~~59.1 &&       0.137 \\
& 88.4 &&        0.139 & ~~88.8 &&      0.141 & ~~90.3 &&    \bf0.102 \\
& 139.1 &&        0.140 & 140.6 &&      0.140 &    136.6 &&       0.147 \\
& 284.7 &&        0.161 & 288.8 &&      0.152 &    282.4 &&       0.156 \\
\midrule
\parbox[t]{2mm}{\multirow{5}{*}{\rotatebox[origin=c]{90}{\cafe}}}
& 31.1 &&     \bf0.112 & ~~32.4 &&      0.131 & ~~31.0 &&       0.114 \\
& 53.5 &&        0.121 & ~~56.0 &&   \bf0.120 & ~~53.7 &&       0.115 \\
& 80.5 &&        0.128 & ~~86.2 &&      0.125 &  ~~84.6 &&     \bf0.109 \\
&   123.7 &&        0.128 &   133.8 &&      0.130 &    130.7 &&       0.123 \\
&   255.9 &&        0.147 &   261.3 &&      0.154 &    261.2 &&       0.170 \\
\bottomrule
\end{tabular} 
}
\begin{tablenotes}[para,flushleft,online,normal] 
{\it Note.} $M$ represents the mean number of tokens in the strata. \\Bold type represents the best score in the each dataset.
\end{tablenotes}
\end{threeparttable}
        \end{center}
\end{table}

\section{Conclusion}
\subsection{Contribution}
In this study, we propose a novel multimodal deep learning model that integrates posted review texts with tabular data, including user profiles and location information. This model effectively captures consumer heterogeneity to predict user ratings on locations with high accuracy. In addition, we conduct a comprehensive analysis of different pre-trained models and the effect of token count on prediction accuracy.

Our proposed model consistently outperforms reference models on test data across all categories. This result indicates the superiority of contextual understanding facilitated by the cross-attention over mere feature fusion for joint representation. Despite prior studies confirming the efficacy of multimodal learning in the various field, in this study, feature fusion which is a simple form of multimodal learning does not overtake of single-modality models. This limitation may stem from the complexity of features, as even with a substantial number of units in the Output-subnet, the large-scale deep-contextualized word representations may overwhelm the upper hidden layers. This result indicates the limitations of simple feature-fusion methods, and as the complexity of the features to be combined reflects, sophisticated mechanisms are needed to understand them.

In addition, our proposed model exclusively utilizes the cross-attention, unlike previous research that emphasizes the combination of features through both attention and feature fusion \cite{alaraj}. Our results demonstrate that achieving higher accuracy is feasible with the cross-attention alone. By establishing causality between different modalities as the source and target, the model can effectively attend to large and sparse features. Although our study focuses on predicting ratings due to data availability, it highlights the potential to construct models based on an accurate understanding of user preferences.

Extending the proposed model presented in this study opens the door to addressing various advanced tasks, such as a model that recommends the appropriate content based on user’s past posts and profile and another model that predicts future repeated purchases based on a consumer's past product reviews and purchase history on the EC platforms.

\subsection{Challenges}
Our model still encounters challenges in improving prediction accuracy, primarily due to computational limitations. All BERT layers in our study remain frozen (i.e., parameters are set to non-trainable) during the training process due to these limitations. In addition, newly developed LLMs are proposed one after another. That is, the model can be further improved through structural refinements, such as selecting different LLMs, fine-tuning BERT layers, incorporating additional dropout, adjusting the number and shape of hidden layers, and optimizing other hyper-parameters.

Finally, despite the use of LLMs, the handling a large number of tokens remains difficult. Our study suggests that an excessive number of tokens may actually decrease prediction accuracy. To address the issue, appropriate measures must be taken, such as pre-summarizing large amounts of text data or using LLMs with larger context windows. It is worth noting that such analyses require additional computational resources and training time, which is a problem to be balanced with prediction accuracy.

\section*{Ethical Statement}
This study only uses academic open data and does not additionally collect personally identifiable information. We observe the terms of use of the dataset and manage the data in a secure environment.

\section*{Acknowledgment}
Our comprehensive analyses were implemented on RAIDEN, a computing infrastructure hosted by RIKEN AIP. We would like to express our gratitude to all the members of the center who maintain the system. Additionally, we extend our gratitude to Yelp which enriched our study by providing the open data.

\bibliographystyle{IEEEtran}  
\bibliography{ieee}

\end{document}